\documentclass[12pt]{article}

\usepackage{amsfonts}
\usepackage{bbm}

\newcommand{\bmat}{\left(\begin{array}}
\newcommand{\emat}{\end{array}\right)}

\def\gtrsim{\mathrel{\raise.3ex\hbox{$>$\kern-.75em\lower1ex\hbox{$\sim$}}}}
\def\NPB#1#2#3{Nucl. Phys. B{#1} (#2) #3}

\def\ts{\textstyle}

\def\a{\alpha}
\def\ap{\alpha^{\prime}}
\def\b{\beta}
\def\g{\gamma}
\def\d{\delta}

\def\vt{\vartheta}

\def\-{\hphantom{-}}
\def\ov{\overline}
\def\s2{\frac{1}{\sqrt2}}

\def\oh{\frac{1}{2}}
\def\beq{\begin{equation}}
\def\eeq{\end{equation}}
\def\beqa{\begin{eqnarray}}
\def\eeqa{\end{eqnarray}}
\def\D{{\rm D}}

\def\re{{\rm Re \,}}

\def\z{\zeta}
\def\T{{\rm T}}
\def\Z{{\mathbb Z}}

\def\cn{{\mathcal N}}

\def\ch{{\cal H}}

\def\mg{m_{3/2}}
\def\mg2{m^2_{3/2}}

\def\deq#1{\mbox{$D$=#1}}
\def\neq#1{\mbox{$\cn$=#1}}
\def\Dsl{\,\raise.15ex\hbox{/}\mkern-13.5mu D} %this one can be subscripted

\newcommand{\mathsmaller}[1]{\mbox{\footnotesize$#1$}}
\begin{document}
\pagestyle{plain}

%----------------------------------------------------------------------%
%  numbering equations with section number
%----------------------------------------------------------------------%
\makeatletter
\@addtoreset{equation}{section}
\makeatother
\renewcommand{\theequation}{\thesection.\arabic{equation}}
%----------------------------------------------------------------------%
%  title page
%----------------------------------------------------------------------%
\pagestyle{empty}
%\vspace*{1.0in}
\rightline{ IFT-UAM/CSIC-04-62}
%\rightline{\tt hep-th/0408036}
\vspace{2.5cm}
\begin{center}
\LARGE{ SUSY-breaking Soft Terms in a MSSM Magnetized D7-brane Model 
\\[10mm]}
\large{A. Font\footnote{On leave from Departamento de F\'{\i}sica, Facultad de Ciencias,
Universidad Central de Venezuela, A.P. 20513, Caracas 1020-A, Venezuela.}
and L.E. Ib\'a\~nez \\[6mm]}
\small{
 Departamento de F\'{\i}sica Te\'orica C-XI
and Instituto de F\'{\i}sica Te\'orica  C-XVI,\\[-0.3em]
Universidad Aut\'onoma de Madrid,
Cantoblanco, 28049 Madrid, Spain 
\\[19mm]} 
\small{\bf Abstract} \\[17mm]
\end{center}

\begin{center}
\begin{minipage}[h]{14.0cm}

We compute the SUSY-breaking soft terms in a magnetized D7-brane model
with MSSM-like spectrum, under the general assumption of non-vanishing
auxiliary fields of the dilaton and K\"ahler moduli. As a particular
scenario we discuss SUSY breaking triggered by ISD or IASD 3-form fluxes.

\end{minipage}
\end{center}
\newpage
%----------------------------------------------------------------------%
%  Resetting of counters
%----------------------------------------------------------------------%
\setcounter{page}{1}
\pagestyle{plain}
\renewcommand{\thefootnote}{\arabic{footnote}}
\setcounter{footnote}{0}
%----------------------------------------------------------------------%
%  Paper begins
%----------------------------------------------------------------------%

%\begin{document}
%\pagestyle{plain}

%\begin{flushleft}
%{\bf Soft Supersymmetry Breaking in Magnetized Brane Models}
%\end{flushleft}

%\vspace*{0.2cm}

\section{Introduction}

Recently a simple intersecting D-brane
model was proposed with massless chiral spectrum close to
that of the MSSM \cite{cim1}. In this model the SM fields lie 
at the intersections of four sets of D6-branes wrapping an
(orientifolded) toroidal compactification of Type IIA 
string theory. The same model may be equivalently described
in terms of different T-dual configurations, e.g. 
in terms of a Type IIB  orientifold with  (magnetized) D9-branes 
and D5-branes \cite{cim2}. Recently \cite{ms} it has been
shown how this type of D-brane configurations may be 
promoted to a fully \neq1 SUSY tadpole-free model (see also \cite{Cvetic:2004nk}) 
by embedding it into a  $\Z_2\times \Z_2$ Type IIB 
orientifold along the lines suggested in \cite{csu}.
A number of results for the effective Lagrangian in such
type of D-brane models is known by now. The Yukawa 
couplings among chiral fields were computed in \cite{cim1,
cvetic,abel,cim2} and other aspects of the effective action
may be found in \cite{Cim1,Cim2,gauge,lmrs,lrs1,jl,lrs2,kn}.
For an up-to-date review, see \cite{blum}. 

One interesting point to address is the structure of
possible SUSY-breaking soft terms in this model.
It has been recently realized that fluxes of antisymmetric 
R-R and NS-NS fields in Type IIB orientifolds may provide a source of 
such terms \cite{Grana, ciu1, ggjl, ciu2, jl,iflux, msw}. It was also 
realized \cite{ciu1} that SUSY breaking from  imaginary self-dual (ISD)
3-form fluxes correspond to a non-vanishing vev for the 
auxiliary field of the overall modulus $T$ and imaginary anti-self-dual (IASD)
correspond to a non vanishing vev for the auxiliary field 
of the complex dilaton $S$. 
On the other hand, a possible phenomenological application of
these  ideas was proposed in \cite{iflux}. In particular, if one 
assumes  that the SM particles correspond
to geometric  D7-brane moduli,  a simple set of SUSY-breaking soft terms 
may be shown to arise from ISD fluxes.

In this article we would like to present explicit results
for the SUSY breaking soft terms in the MSSM-like model of 
ref.~\cite{cim1} as a function of the vevs of the auxiliary fields
of the K\"ahler moduli $T_i$  and/or the complex dilaton $S$. As particular 
examples we consider vevs induced by ISD and/or IASD 3-form fluxes.
We  construct the model \cite{cim1} in
terms of 3 stacks of intersecting D7-branes, one of them containing
a constant magnetic field (leading to chirality and family replication).
We then use the effective supergravity Lagrangian approach in
order to obtain soft terms, as in ref.~\cite{soft,bim2}.

A previous detailed analysis of these soft terms 
including the effect of the non-vanishing magnetic flux 
was presented by L\"ust, Reffert and Stieberger in ref.~\cite{lrs2}, 
based on the K\"ahler metrics of matter fields
computed in ref.~\cite{lmrs, lrs1}. These included the 
effect of magnetic fluxes. Some phenomenological analysis of those
results was described in \cite{kane}. Soft terms in the T-dominance case 
were briefly discussed in \cite{iflux} following \cite{imr}, in which
the effect of magnetic fluxes was not included. 
In the present paper we revisit previous results taking into account 
a proper normalization of the matter fields. We also
make use of the K\"ahler metrics for chiral fields
discussed in ref.~\cite{lmrs, lrs1}. 
Including a factor implied by the analysis of \cite{lmrs}
the SUSY-breaking soft terms simplify considerably. 
One of the motivations of the present work was to find the
connection with the analogous results obtained in \cite{imr}
in the absence of magnetic fluxes. Indeed, we find that 
in the limit of diluted magnetic fluxes those results are recovered.  

It is known that NS-NS and R-R fluxes on {\em toroidal } settings 
induce soft terms on D3-brane fields of order $M_s^2/M_{Pl}$, so that  
one can obtain a hierarchy of scales by lowering the string scale \cite{ciu1}.   
However, in the case of intersecting D7-branes, 
as in the model at hand, one cannot lower the string scale without 
making the SM gauge couplings unacceptably small.
Therefore, in toroidal/orbifold models with intersecting D7-branes 
the fluxed-induced soft terms are typically of order the string scale.
This fact is due to the simplicity of toroidal compactifications
in which the compact space is flat and the fluxes are distributed uniformly.
In a generic Calabi-Yau (CY) compactification this is
not going to be the case and there may be regions in the CY in which
fluxes are concentrated and others in which fluxes are diluted.
This possibility was considered e.g. in \cite{gkp, kklt} in order
to obtain hierarchies. Thus, for generic CY compactifications 
the size of soft terms will actually depend on the detailed geometry of the 
fluxes in the CY.  

The local set of branes leading to a MSSM-like spectrum introduced in \cite{cim1} is 
nevertheless likely to be more generic than the toroidal setting in which it was
first proposed (see e.g. \cite{bbkl}).
In particular, it has recently been shown \cite{schellekens}
that there are many thousands of models with the 4-stacks of
branes structure of the model in \cite{cim1} (these are labeled Type-4
models in ref.~\cite{schellekens}).  Therefore, one may expect to obtain 
this MSSM structure in CY orientifold models beyond the toroidal 
setting. In these more general models the size of induced soft terms
may not be tied to the string scale and could be much lower. 
In our effective field theory analysis below we will not commit 
ourselves to a particular scale for the soft terms. Instead,
following \cite{soft,bim2}, we will assume that the effect of SUSY-breaking
is encoded in non-vanishing vevs for the auxiliary fields of
the complex dilaton and K\"ahler moduli. However, we also
discuss the case in which the source of SUSY-breaking 
are constant IASD or/and IASD 3-form fluxes, which correspond to a 
particular choice for the auxiliary fields.

\section{A MSSM-like model from magnetized D7-branes}
\label{sec:setup}

We will construct the model in \cite{cim1} in terms of 
three sets of intersecting Type IIB D7-branes (see e.g.\cite{ms}).
We consider type IIB string theory compactified on a factorized 
six-torus $\T^6= \otimes_{i=1}^{3} \T^2_i$. 
We will further do an orientifold projection by 
 $\Omega (-1)^F I_6$, $\Omega $ being the world-sheet parity operator 
and $I_6$ a simultaneous reflexion of the six toroidal coordinates.
We also include sets of ${\rm D}9_a$-branes and allow for
possible constant magnetic fluxes across any of the three 2-tori
\beq
\frac{m_a^i}{2\pi} \int_{\T^2_i} \, F_a^i = n_a^i  \ ,
\label{nmdef}
\eeq
where $F_a^i$ is the world-volume magnetic field.
For each group of branes the
state of magnetization is thus characterized by the integers $(n_a^i, m_a^i)$,
where $m_a^i$ is the wrapping number and $n_a^i$ is the total magnetic
flux. It is useful to introduce the angles
\beq
\psi_a^i = \arctan 2\pi\ap F_a^i= \arctan \frac{\ap n_a^i}{m_a^i A_i}  \,
\label{psidef}
\eeq
where $(2\pi)^2A_i$ is the area of the $\T^2_i$.
The magnetized
${\rm D}9_a$-branes \cite{bachas, bgkl, aads} preserve the same supersymmetry of 
the orientifold planes provided that \cite{blt, cu}
\beq
\sum_{i=1}^{3} \psi_a^i = \frac{3\pi}2 \, {\rm mod} \, 2\pi  \ .
\label{susycon}
\eeq
Note that in this scheme 
lower dimensional branes are described setting $m_a^i=0$ for all $i$ transverse
to the brane. For example, a D3-brane has  $(n_a^i,m_a^i)=(1,0)$, $i=1,2,3$. 
Notice in particular that the D3-brane satisfies (\ref{susycon}).
T-duality along the horizontal direction in each $\T^2_i$ gives the dual
picture of D6-branes at angles. For example, for square $\T^2_i$, $A_i=R_{ix} R_{iy}$,
and the dual angle is $\vartheta_a^i = \arctan (n_a^i R_{ix}/m_a^i R_{iy})$.

In order to reproduce the structure of the MSSM-like model of ref.~\cite{cim1,cim2}
one introduces three sets of $\D7_i$-branes $i=1,2,3$  which are characterized by 
being transverse to the $i$-th 2-torus. In particular the relevant magnetic data is
\beq
\begin{tabular}{ccccc}
Branes & $(n_a^1,m_a^1)$ & $(n_a^2,m_a^2)$ & $(n_a^3,m_a^3)$ & 
$(\psi_a^1, \psi_a^2, \psi_a^3)$ \\[0.2ex]
${\rm D}7_1$ & (1,0) & $(g,1)$ & $(g,-1)$ & 
$(\displaystyle{\frac{\pi}2}, \pi\d_2, \pi -\pi\d_3)$ \\
${\rm D}7_2$ & (0,1) & $(1,0)$ & $(0,-1)$ & $(0, \displaystyle{\frac{\pi}2}, \pi)$ \\
${\rm D}7_3$ & (0,1) & $(0,-1)$ & $(1,0)$ & 
$(0, \pi, \displaystyle{\frac{\pi}2})$   
\end{tabular}§  
\ \ .
\label{sbs}
\eeq
where $\pi\d_i=\arctan (\ap g/A_i)$. 
We will take $\D7_1$-branes to  come in four copies so that generically the
associated gauge group will be $U(4)$. Branes $\D7_2$ and $\D7_3$ 
come only in one copy and are located on top of the orientifold 
plane at the origin so that they give rise to a
gauge group $Sp(2)\times Sp(2)\simeq SU(2)\times SU(2)$.
Altogether the overall gauge group is $U(4)\times SU(2)\times SU(2)$.
It may be shown that the $U(1)$ 
(which corresponds to $(3B+L)$) is anomalous and becomes massive in the
usual way by the St\"uckelberg mechanism. In fact, one can further
make the breakings $SU(4)\rightarrow SU(3)_c\times U(1)_{B-L}$ 
and $SU(2)_R\rightarrow U(1)_R$ by
e.g. Wilson lines. Thus the final gauge group is just
$SU(3)\times SU(2)\times U(1)_Y\times U(1)_{B-L}$. 
The final chiral spectrum is displayed in Table 1 
for the choice $g=3$ which leads to three quark/lepton generations.
\begin{table}[htb] \footnotesize
\renewcommand{\arraystretch}{1.25}
\begin{center}
\begin{tabular}{|c|c|c|c|c|}
\hline Intersection &
 Matter fields  & Rep.   &  $Q_{B-L}$     & Y \\
\hline\hline  $\D7_1-\D7_2$  & $Q_L$ &  $3(3,2)$ &   1  & 1/6 \\
\hline  $\D7_1-\D7_3$  & $U_R$ &  $3({\bar 3},1)$ & -1   &  -2/3 \\
\hline  $\D7_1-\D7_3$  & $D_R$ &  $3({\bar 3},1)$ & -1   &  1/3 \\
\hline  $\D7_1-\D7_2$  & $E_L$ &  $3(1,2)$ & -1   &  1/2 \\
\hline  $\D7_1-\D7_3$  & $E_R$ &  $3(1,1)$ & 1   &  -1 \\
\hline  $\D7_1-\D7_3$  & $N_R$ &  $3(1,1)$ & 1   &  0 \\
\hline  $\D7_2-\D7_3$  & $H$ &  $(1,2)$ & 0   &   1/2 \\
\hline  $\D7_2-\D7_3$  & ${\bar H}$ &  $(1,2)$ & 0   &   -1/2 \\
\hline \end{tabular}
\end{center} \caption{\small  Chiral spectrum of the MSSM-like model.}
\label{mssm}
\end{table}

The ${\rm D}7_2$ and ${\rm D}7_3$ do not have magnetic flux and do verify 
the supersymmetric condition (\ref{susycon}). 
However, for the case of the ${\rm D}7_1$-branes, with opposite magnetic fields
turned on in the second and third $\T^2$, to be supersymmetric we
need to impose 
\beq
A_2=A_3=A \ , 
\label{susya}
\eeq
so that $\d_2=\d_3=\d$ and
\beq
\tan \pi \d = \frac{\ap g}{A} .
\label{deldef}
\eeq
Clearly, the condition (\ref{susycon}) also guarantees that any two 
sets of branes preserve a common supersymmetry. Notice that the relative angles 
\beq
\theta^i_{ab} = \psi^i_b - \psi_a^i 
\label{reltheta}
\eeq
automatically satisfy $\sum_{i} \theta^i_{ab} = 0 \, {\rm ºmod} \, 2\pi$. 

Departures from the equality $A_2=A_3$ may be shown \cite{cvetic, Cim1, Cim3, cgqu}
to correspond to a non-vanishing Fayet-Iliopoulos term for the anomalous
$U(1)_{3B+L}$. In fact, for $A_2=(A_3\ +\ \epsilon)$ 
and small $\epsilon $ one finds \cite{Cim1, Cim3}
\beq
\xi _{FI} \ =\ \frac {g\epsilon}{A^2 + \a^{\prime \, 2} g^2}
\label{fi}
\eeq
where the D-term potential is of the form
\beq
V_{FI}(\phi_n) = {1 \over 2 g_{U(1)}^2} (\sum_n q_n|\phi_n|^2 + \xi_{FI})^2.
\label{potentialFI2}
\eeq
and $\phi_n$ runs over squarks and sleptons.
Left-handed and right-handed chiral fields have positive
and negative  $U(1)_{3B+L}$
charge respectively so that a non-vanishing $\epsilon$  may
induce further symmetry breaking. 
Note that this potential as it stands does not prefer $\xi_a = 0$
(and hence the SUSY condition $A_2=A_3$)
as sometimes claimed in the literature, since a non-vanishing 
$\xi_a $ may always be compensated with a vev for a right-handed
scalar field (e.g. the right-handed sneutrino).
On the other hand, in the presence 
of soft masses for the chiral fields, as shown to appear in the next sections,
a vanishing FI-term is dynamically preferred and so is the SUSY condition
$A_2=A_3$. As we will see, this implies in turn the unification of
$SU(2)_L$ and $SU(2)_R$ gauge couplings.

Let us finally comment that, as it stands, this brane configuration 
has R-R tadpoles so some additional (`hidden') brane system should be
added. This can be done in a way consistent with \neq1 SUSY 
if we embed this brane configuration in a $\Z_2\times \Z_2$ orientifold
\cite{csu} as recently shown in \cite{ms} (although in this case
one cannot do the breaking $SU(2)_R\rightarrow U(1)_R$ via Wilson lines \cite{ms}).
 Since we are only
interested in
the structure of soft terms for the MSSM fields we will not deal here
with these global issues of the compactification. Our results will still hold
for those global generalizations.

\section{Massless fields and effective supergravity action}
\label{sec:action}
 
Let us now turn to the effective supergravity action in this model.
We will compile general formulas for the K\"ahler potential, matter 
metrics and gauge kinetic function and we will apply them to the 
specific D-brane model at hand.

We begin with the field content.  
In the closed string sector, in addition to the supergravity multiplet, 
one has   the dilaton $S$ plus the K\"ahler and complex structure moduli. 
We use conventions such that the complex dilaton is given by 
\beq
S= e^{-\phi_{10}} + i a_0   \ ,
\label{sdef}
\eeq
where $a_0$ is the R-R 0-form. Recall that the string coupling constant is 
$g_s=e^{\phi_{10}}$.

For the metric moduli we will restrict for simplicity here to the
diagonal fields $U_j$, $T_j$, $j=1,2,3$. Note that in any case 
the off-diagonal fields would not be
present in a  $\Z_2\times \Z_2$ embedding
of the present model.
To be more concrete,
let the $\T^2_j$ lattice vectors be denoted $e_{jx}$, $e_{jy}$. Then, the
geometric toroidal moduli are
\beqa
\tau_j & = & \frac1{e^2_{jx}}(A_j + i\,  e_{jx} \cdot e_{jy}) 
\nonumber \\[0.2cm]
\rho_j & = &  A_j + i a_j \  ,
\label{gmod}
\eeqa
where the axions $a_j$ arise from the R-R 4-form. In type IIB, the 
toroidal complex structure $\tau_j$ is equal to the moduli field
$U_j$ that appears in the \deq4 supergravity action. However,
the correct K\"ahler moduli field $T_j$ is not $\rho_j$.
One way to see this is to realize that  the gauge coupling squared 
of {\em unmagnetized} ${\rm D}7_j$-branes should be equal to
 $2\pi/\re T_j$.  Then, since, e.g. 
a ${\rm D}7_1$ wraps $\T^2_2$ and  $\T^2_3$, from the Born-Infeld
action it follows that $\re T_1 =  e^{-\phi_{10}} A_2 A_3/\a^{\prime \, 2}$.
In general
\beq
T_i = e^{-\phi_{10}} \frac{A_j A_k}{\a^{\prime \, 2}} + i a_i  \quad ; \quad
j\not=k\not=i   \ .
\label{tdef}
\eeq
For later convenience we define
\beq
s=S+ \ov{S}  \quad ; \quad   t_i=T_i + \ov{T}_i \quad ; \quad
u_i=U_i + \ov{U}_i    \   .
\label{minimod}
\eeq
The \deq4 gravitational coupling is $G_N=\kappa^2/8\pi$ where 
\beq
\kappa^{-2} = \frac{M_{Pl}^2}{8\pi} =  e^{-2\phi_{10}} 
\frac{A_1 A_2 A_3}{\pi\a^{\prime \, 4}} \ 
=  \frac {(st_1t_2t_3)^{1/2}} {4\pi \ap}   \   .
\label{mplanck}
\eeq
It is also useful to introduce the T-duality invariant four-dimensional dilaton, namely
\beq
\phi_4 = \phi_{10} - \oh \log(A_1 A_2 A_3/\a^{\prime\, 3}) \  .
\label{phi4}
\eeq
Notice that $\kappa^{-2} =  e^{-2\phi_4}/\pi \ap$.
The string scale is $M_s=1/\sqrt{\ap}$.
%The coefficient of the Einstein term in the supergravity action is $M_{Pl}^2/16\pi$. 

Open strings give rise to charged fields. 
We call `untwisted' the states corresponding to
open strings beginning and ending on the same stack of branes, whereas
`twisted' refers to the chiral fields lying at the intersection of two
different stacks of D7-branes. For the content of
branes in (\ref{sbs}), and assuming supersymmetry is preserved, the untwisted 
sectors  ${\rm D}7_i$-${\rm D}7_i$, $i=1,2,3$,  give a gauge multiplet of a 
group $G_i$ and 3 massless chiral multiplets, denoted $C_j^{7_i}$, $j=1,2,3$,
transforming in the adjoint of $G_i$. The $C_j^{7_i}$ are the ${\rm D}7_i$-brane
moduli, $C_i^{7_i}$ gives the position of the brane in the transverse $\T^2_i$
whereas $C_j^{7_i}$, $j\not= i$, correspond to Wilson lines
on the two internal complex dimensions parallel to the $\D7_i$-brane.
{} From the twisted sectors  ${\rm D}7_i$-${\rm D}7_j$
there are only chiral massless multiplets, denoted $C^{7_i 7_j}$, transforming as
bifundamentals of $G_i \times G_j$. 

The low-energy dynamics of the massless fields is governed by a \deq4, \neq1
supergravity action that depends on the K\"ahler potential, the gauge kinetic
functions and the superpotential. In particular, the F-part of the scalar
potential is
\beq
V = e^{\kappa^2 K} \left[ K^{\bar A B}  (D_AW)^* (D_BW) - 3 \kappa^2 |W|^2 \right] \ ,
\label{sugrapot}
\eeq 
where $D_A W = \partial_A W + \kappa^2 \partial_A K W$ and
$K^{\bar A B}$ is the inverse of $K_{\bar A B}= \partial_{\bar A} \partial_B K$.
Recall that the auxiliary field of a chiral superfield $\Phi_A$ is
\beq
\bar F^{\bar A} = \kappa^2 e^{\kappa^2 K/2} K^{\bar A B}  D_BW  \ .
\label{fterm}
\eeq
We now describe the functions $K(\phi; \bar \phi)$,
$f_i(\phi)$ and $W(\phi)$ in our setup, in which $\Phi_A = \{M, C_I\}$, with 
$M=\{S, T_i, U_i\}$ and  $C_I=\{C_j^{7_i}, C^{7_i 7_j} \}$.

%\vspace*{0.2cm}

%\noindent
%\begin{flushleft}
%\underline{K\"ahler potential}
%\end{flushleft}

%\noindent
\subsection{K\"ahler potential}

The K\"ahler potential has the structure
\beq
K = \hat K(M,\bar M) + \sum_{I,J} \tilde K_{I \bar J}(M,\bar M)  C_I \bar C_J  
+ \oh \sum_{I,J} [ Z_{IJ}(M,\bar M)  C_I C_J + c.c.] + \cdots   \ .
\label{kpot}
\eeq
The contribution of the closed string moduli is
\beq
\kappa^2 \hat K(M,\bar M) = -\log s  -\sum_{i=1}^3 \log t_i -\sum_{i=1}^3 \log u_i
\ .
\label{kmodu}
\eeq
Below we describe in detail the K\"ahler metrics of matter fields.

The $\tilde K_{I \bar J}$ for unmagnetized branes were deduced
in \cite{imr} using T-duality arguments. For generic magnetized branes they have been
obtained in \cite{lmrs, lrs1} {}from a computation of string scattering amplitudes.
These metrics vanish when $J \not= I$. 
Below we present the diagonal entries for all possible cases with the brane
content of (\ref{sbs}). To streamline notation 
we write $\tilde K_{i,j \bar \jmath}= \tilde K_{C_j^{7_i} \bar C_j^{7_i}}$ 
in untwisted sectors, and  $\tilde K_{ij,C \bar C}= 
\tilde K_{C^{7_i 7_j} \bar C^{7_i 7_j}}$
in twisted sectors. For untwisted fields we have

\begin{trivlist}

\item[$\bullet$]
${\rm D}7_3$-${\rm D}7_3$ (untwisted, unmagnetized)
\beq
\kappa^2 \tilde K_{3,1 \bar 1}  =  \frac1{u_1 t_2} 
\quad ; \quad
\kappa^2 \tilde K_{3,2 \bar 2}  =  \frac1{u_2 t_1}
\quad ; \quad
\kappa^2 \tilde K_{3,3 \bar 3}  =  \frac1{u_3 s}   \  .
\label{stst} 
\eeq

\item[$\bullet$]
${\rm D}7_2$-${\rm D}7_2$ (untwisted, unmagnetized)
\beq
\kappa^2 \tilde K_{2,1 \bar 1}  =  \frac1{u_1 t_3}
\quad ; \quad
\kappa^2 \tilde K_{2,2 \bar 2}  =  \frac1{u_2 s}
\quad ; \quad
\kappa^2 \tilde K_{2,3 \bar 3}  =  \frac1{u_3 t_1}   \  .
\label{sdsd} 
\eeq

\item[$\bullet$]
${\rm D}7_1$-${\rm D}7_1$ (untwisted, magnetized)
\beq
\kappa^2 \tilde K_{1,1 \bar 1}  =  \frac1{u_1 t_1 s}(g^2 s + t_1)
\quad ; \quad
\kappa^2 \tilde K_{1,2 \bar 2}  =  \frac1{u_2 t_2}
\quad ; \quad
\kappa^2 \tilde K_{1,3 \bar 3}  =  \frac1{u_3 t_3}  \  .
\label{susu} 
\eeq
To obtain these results we start from the general expressions
in the geometric basis given in \cite{lrs1}. In our notation these are
\beqa
\kappa^2 \tilde K_{1,1 \bar 1} & = & \frac{e^{\phi_4}}{\a^{\prime\, 2}\, u_1}
\sqrt{\frac{\ap A_1}{A_2 A_3}} |A_3 m_1^3 + i \ap n_1^3| |A_2 m_1^2 + i \ap n_1^2| \  ,
\nonumber \\[0.2cm]
\kappa^2 \tilde K_{1,2 \bar 2} & = & \frac{e^{\phi_4}}{u_2}
\sqrt{\frac{\ap A_2}{A_1 A_3}} \left|\frac{ A_3 m_1^3 + 
i \ap n_1^3}{A_2 m_1^2 + i \ap n_1^2} \right|   \  ,
\label{sususb} \\[0.2cm]
\kappa^2 \tilde K_{1,3 \bar 3} & = & \frac{e^{\phi_4}}{u_3}
\sqrt{\frac{\ap A_3}{A_1 A_2}} 
\left|\frac{ A_2 m_1^2 + i \ap n_1^2}{A_3 m_1^3 + i \ap n_1^3} \right|
\  .
\nonumber
\eeqa
We then substitute the values of the $(n_1^i, m_1^i)$ given in (\ref{sbs}), use
the supersymmetry condition (\ref{susya})  and also
\beq
\a^{\prime \, 2} t_1 = s A^2 
\label{agf}
\eeq
that follows from (\ref{tdef}). When $m_1^i=0$, (\ref{sususb}) gives the metric
of a ${\rm D}3$-${\rm D}3$ sector.  When $n_1^i=0$, we just obtain the metric
of unmagnetized ${\rm D}7_1$-${\rm D}7_1$.

\end{trivlist}

For the metrics of twisted fields one has

\begin{trivlist}

\item[$\bullet$]
${\rm D}7_2$-${\rm D}7_3$ (twisted, unmagnetized)
\beq
\kappa^2 \tilde K_{23,C \bar C}  =  \frac1{(u_2 u_3 s t_1)^{1/2}}
\label{sdst} \  .
\eeq

\item[$\bullet$]
${\rm D}7_1$-${\rm D}7_2$ (twisted, magnetized) 
\beq
\kappa^2 \tilde K_{12,C \bar C}  =  \frac1{(s t u_1)^{1/2} u_2^{1/2 + \d} u_3^{1-\d}}
\, \frac{\Gamma(\oh - \d)}{\Gamma(1-\d)}  \  ,
\label{susd} 
\eeq
where $t=t_2=t_3$. Notice that $\d$ depends implicitly on $s$ and $t_1$. {}From
(\ref{deldef}) and  (\ref{agf}),
\beq
\tan \pi\d = g (s/t_1)^{1/2} \  .
\label{tanpid}
\eeq
Observe that $0 \leq \d < \oh$. To derive (\ref{susd}) we start from 
\beq
\kappa^2 \tilde K_{12,C \bar C}  =   e^{\phi_4}
\, \prod_{j=1}^3 u_j^{-\nu_j} \, \sqrt{\frac{\Gamma(1-\nu_j)}{\Gamma(\nu_j)}} \ ,
\label{kori} 
\eeq
where the $\nu_j$, computed from $\hat \nu_j= \theta^j_{12}/\pi$, are such that
$0 \leq \nu_j < 1$ and $\nu_1 + \nu_2 + \nu_3 = 2$. 
To determine the $\nu_j$, observe first that $\hat \nu_1 + \hat \nu_2 + \hat \nu_3 = 0$.
Assuming $\hat \nu_j \not=0$ then implies that one or two of the $\hat \nu_j$ are negative.
In the first case start instead from $\hat \nu_j= \theta^j_{21}/\pi$. Now two of
the $\hat \nu_j$ are negative by construction. Finally, define
$\nu_j=1 + \hat \nu_j$ if $\hat \nu_j$ is negative, otherwise $\nu_j=\hat \nu_j$. 
In this case $\nu=(\oh, \oh + \d, 1-\d)$.
 
To arrive at (\ref{susd}) we use (\ref{tanpid}) and the relation
\beq
\frac{\Gamma(\d)}{\Gamma(\oh+\d)} = 
\frac{\Gamma(\oh - \d)}{\tan \pi\d \, \Gamma(1-\d)}  \ . 
\label{gprop}
\eeq
In (\ref{susd}) we can take the limit $\d \to 0$ and recover the metric of 
unmagnetized ${\rm D}7_1$-${\rm D}7_2$, provided we drop $u_3$ that 
would have exponent -1.
Using again (\ref{tanpid}) and (\ref{gprop}) we can also take the limit $\d \to \oh$ 
and, dropping $u_2$ now with exponent -1, retrieve the metric of ${\rm D}3$-${\rm D}7_2$, 
namely $\kappa^2 \tilde K_{C \bar C} = (t_1 t_3 u_1 u_3)^{-1/2}$. 
The fact that moduli $u_j$ with would be exponent -1 do not appear in the metric
also occurs in twisted sectors of heterotic orbifolds \cite{dkl}.

Eq. (\ref{kori}), including the prefactor $e^{\phi_4}$, follows putting together 
results found in \cite{lmrs}.
In the field basis this prefactor can be recast as $2(s t_1 t_2 t_3)^{-1/4}$. 
The square root of Gamma functions, with the arguments as shown in (\ref{kori}),
is determined by the differential equation that dictates the dependence 
on the K\"ahler moduli \cite{lmrs}. 
Finally, for the remaining twisted sector the metric is

\item[$\bullet$]
${\rm D}7_1$-${\rm D}7_3$ (twisted, magnetized) 
\beq
\kappa^2 \tilde K_{13,C \bar C}  =  \frac1{(s t u_1)^{1/2} u_2^{1-\d} \, u_3^{1/2+ \d}}
\, \frac{\Gamma(\oh - \d)}{\Gamma(1-\d)}  \  .
\label{sust}
\eeq

\end{trivlist}

\subsection{Gauge kinetic functions}

The gauge kinetic functions $f_i$ for the groups arising 
in the ${\rm D}7_i$-${\rm D}7_i$ 
sectors are
\beq
f_1  =  T_1 + g^2 S \quad ; \quad
f_2  =  T_2 \quad ; \quad
f_3= T_3  \  .
\label{efes}
\eeq
In general \cite{Cim1, lmrs},
\beq
\re f_i = \frac{e^{-\phi_{10}}}{\a^{\prime \, 2}} \, 
\prod_{j \not= i} |m_i^j A_j + i\ap n_i^j|  \  .
\label{realfgen}
\eeq
Substituting the values of the $(n_i^j, m_i^j)$ given in (\ref{sbs}) leads
to (\ref{efes}). 

Note that if the SUSY condition (\ref{susya}) is verified,
one has $\re f_2= \re f_3$ and the $SU(2)_L$ and $SU(2)_R$ gauge couplings 
are unified. Note also that if the complete model has additional 
branes (as in e.g., \cite{ms})
the SUSY conditions may involve in general also the  area $A_1$ of 
the first torus and imply further unification constraints. 

Concerning the axions, one can check that the linear combination 
$(a_2-a_3)$ becomes massive combining with the anomalous $U(1)_{3B+L}$ gauge
boson through the St\"uckelberg mechanism. On the other hand, one can
also check that the linear combination $(9a_0-a_1)$ has
axionic couplings with the QCD gauge bosons. This may help in
solving the strong CP problem.

Further inspection of the $f_i$ reveals an interesting bound on the string
coupling constant in the D-brane model. {}From (\ref{tanpid}) and the
$SU(4)$ gauge coupling $\a_1$ we deduce the relation
\beq
\sin^2 \pi\d = \frac{2 \a_1 g^2}{g_s}  \  .  
\label{gsbound}
\eeq
This has a number of consequences. 
To have three generations, $g=3$. Hence, the above relation implies
$\a_1 \leq g_s/18$. For $\a_1(M_s) \sim 1/24$ this is consistent with
$g_s < 1$. However, we already see that to get values of coupling
constants of the order of (extrapolated) known gauge couplings, the
string coupling constant $g_s$ approaches the non-perturbative regime. 
In fact, in the present specific toroidal model in addition to the chiral
spectrum there are massless chiral adjoints which will make the 
gauge interactions asymptotically non-free. Thus, the $\alpha_i$ will be
larger than  $\sim 1/24$ and hence the above bound will give $g_s >O(1)$. 
A second (related) implication of eq.(\ref{gsbound}) is that in order to
accommodate gauge couplings consistent with experiment but still stay
within the string perturbative regime with $g_s<1$, the value of  $\delta$
(and hence the magnetic flux) will be substantial. Thus, e.g. for
$\alpha _1=1/24$ one finds $\pi\delta \sim 60^0$. 
All this tells us that in order to have consistency with
the observed values of gauge couplings in this class of intersecting
D7-brane models we would probably need to go to a non-perturbative
F-theory (or perhaps simply non-toroidal) versions  
of them. Constructing such class of F-theory models
would be a rather non-trivial task.  
In what follows we will not further deal with these
issues and simply assume that we still 
remain in a perturbative regime, hoping that a proper fit
of experimentally measured gauge couplings does not substantially
modify our soft term results.

\subsection{Superpotential}

The superpotential can be written as
\beq
W(M, C_I) = \hat W(M) + \oh \sum_{I,J} \mu_{IJ}(M) C_I C_J + 
\frac16  \sum_{I,J,L} Y_{IJL}(M) C_I C_J C_L + \cdots    \ .
\label{superw}
\eeq
For the brane content of (\ref{sbs}), the cubic couplings allowed
are of the form \cite{bl}
\beq
C_1^{7_i} C_2^{7_i} C_3^{7_i}  \quad ; \quad
d_{ijk} C_i^{7_j} C^{7_j 7_k} C^{7_j 7_k}  \quad ; \quad
 C^{7_1 7_2} C^{7_2 7_3} C^{7_3 7_1} \  ,
\label{cubic}
\eeq
where $d_{ijk}=1$ if $i\not=j\not=k$, otherwise $d_{ijk}=0$.
Note that for the case at hand the first type of couplings in
(\ref{cubic}) correspond to standard \neq4 Yukawa couplings among
adjoints. Concerning the second type of couplings, only
one of type $XH{\bar H}$ is present in the model.
Here $X$ is a linear combination of $C_1^2,C_1^3$, the
latter corresponding to Wilson line chiral fields of the 
branes $\D7_2$, $\D7_3$ in the first complex plane.
Note that a vev for $X$ would render the Higgs multiplets massive
so $X$ behaves as a $\mu $-term in the effective Lagrangian\footnote{In the  
T-dual version in terms of intersecting D6-branes, $\langle X \rangle$ 
corresponds to the distance between the $SU(2)_L$ and $SU(2)_R$ D6-branes 
in the first complex plane.}. 
Finally, the third type of superpotential couplings corresponds to the 
regular Yukawa couplings between the chiral generations and the Higgs multiplets.
All in all, the perturbative superpotential among the chiral open 
string multiplets in this model has the general expression
(we write it for the extended gauge group $SU(4)\times SU(2)_L\times SU(2)_R$
for simplicity of notation)
\beq
W_{Yukawa}\ =\ \sum_{i}C_1^{7_i} C_2^{7_i} C_3^{7_i} \ +\
X \ch \ch \ +\
\sum _{\a,\b} h_{\a\b} \ch L_\a R_\b   \  ,
\label{yukis}
\eeq
where $\ch$, $L_\a$ and $R_\b$ are the Higgs and chiral fermions transforming as
$(1,2,2)$, $(4,2,1)$ and $({\bar 4},1,2)$ respectively ($\a,\b=1,2,3$ are generation labels).

The superpotential  couplings $h_{\a\b}$ for this toroidal model
have been computed in \cite{cim1,cim2} and are given by
\beq
h_{\a\b} \ =\ 
 \vt  
\left[{\frac{\a}3 \atop \mathsmaller{0}} \right]
\left(3\z_2, 3U_2\right)
\vt
\left[{\frac{\b}3 \atop \mathsmaller{0}} \right]
\left(3{\z}_3, 3U_3  \right)    \   ,
\label{ahuevo}
\eeq
where $\z_i$ are certain combinations of singlet $C_i^{7_j}$ fields
(see \cite{cim1,cim2}) and $\vt $ are Jacobi theta functions.
For our purposes the only relevant thing to point out is that 
these superpotential couplings only depend on the complex structure 
moduli $U_2$, $U_3$, and not on the K\"ahler moduli nor the dilaton.

In principle we can use the above results to compute 
soft terms under the general assumption that the auxiliary fields 
of the moduli and dilaton are non-vanishing, 
in the spirit of refs.\cite{soft,bim2}. We would not need then to
specify the microscopic source of this values. On the other hand, lately we have
learned that such microscopic source of SUSY-breaking may be provided by 
fluxes in Type II theory.
Hence in addition to the above perturbative chiral couplings, a
moduli dependent superpotential $\hat W(M)$ may be present.
In particular, it is known that
antisymmetric R-R and NS-NS fluxes $F_3,H_3$ generate
a superpotential \cite{gvw}
\beq
W_f = \int G_3 \wedge \Omega    \   
\label{wflux}
\eeq
that depends on $S$ and the $U_i$.
Here $G_3 = F_3 - i S H_3$ and  $\Omega$ is the holomorphic (3,0) form of $\T^6$.
Besides, there may be non-perturbative interactions (like e.g. those from gaugino 
condensation) giving rise to a generic superpotential $W_{np}$. Thus the total
moduli-dependent superpotential will have the general form
\beq
\hat W(M) = W_f(S,U_i) + W_{np}(S,U_i,T_i)  \  .
\label{wmodu}
\eeq
In the absence of $W_{np}$
the equations of motion require $G_3$ to be imaginary self-dual
(ISD), meaning that $G_3$ is a combination of (0,3) and (2,1) fluxes \cite{gkp}.
In this case $D_S W_f =0$ and $D_{U_i} W_f =0$ but $D_{T_i} W_f \not=0$ because
$W_f$ does not depend on the $T_i$ and a (0,3) piece in $G_3$ generates
$W_f \not= 0$.

\section{Soft Terms}
\label{sec:soft}

Armed with all the above data for the low-energy effective action
we can now compute the SUSY-breaking soft terms. To this purpose 
we will follow the approach in \cite{soft,bim2} and assume that the
auxiliary fields $F_{T_i},F_S$ of the  K\"ahler moduli and the
complex dilaton acquire non-vanishing expectation values.
We will later consider the particular case in which 
the microscopic origin of such non-vanishing values is 
provided by ISD and IASD three-form fluxes.
The standard results for the normalized soft parameters 
may be found e.g. in \cite{bim2} and read
\beqa
M_i & = & \frac1{2 \re f_i} F^M \partial_M f_i   \  ,
\nonumber \\[0.2cm]
m^2_I & = & m^2_{3/2} + V_0 - \sum_{M,N} \bar F^{\bar M} F^N 
\partial_{\bar M} \partial_N \log(\tilde K_{I \bar I})   \  ,
\label{softt} \\
A_{IJL} & = & F^M [ \hat K_M  + \partial_M \log(Y_{IJL}) - 
\partial_M \log(\tilde K_{I \bar I} \tilde K_{J \bar J} \tilde K_{L \bar L}) ]  \  .
\nonumber
\eeqa
Here $V_0$ is the vev of the scalar potential and the gravitino mass is
\beq
m_{3/2}= e^{\kappa^2 K/2} |W| \ .
\label{mgravi}
\eeq
Note that, as pointed out above, the $Y_{IJL}$ superpotential
couplings  do not depend on the $S$ and $T_i$ fields,
so that the second contribution to
$A_{IJL}$  in (\ref{softt}) vanishes identically.
The expressions (\ref{softt}) are 
valid when $\tilde K_{I \bar J} \propto \delta_{I\bar J}$ 
which is our case.

The vevs of the auxiliary fields are conveniently parametrized as \cite{bim2}
\beqa
F^S & = & \sqrt3 s C m_{3/2} \sin \theta e^{-i\g_S} 
\nonumber \\[0.1cm]
F^{T_i} & = & \sqrt3 t_i  \eta_i C m_{3/2} \cos \theta e^{-i\g_i} \   ,
\label{vevstino}
\eeqa 
where the goldstino angle $\theta$ and the $\eta_i$, with $\sum_i \eta_i^2 =1$,
control whether $S$ or the $T_i$ dominate SUSY breaking. We further assume that
$F^{U_i}=0$. Then, substituting in (\ref{sugrapot}) gives
\beq
C^2= 1 + \frac{V_0}{3  m_{3/2}^2} \  .
\label{cconst}
\eeq 
We will now present the soft terms for the D-brane model.

\noindent
\begin{flushleft}
\underline{Gaugino masses}
\end{flushleft}

\noindent
Masses for gauginos of the group $G_i$ arising in the $\D7_i$-$\D7_i$ sector
are denoted $M_i$ or $M_{G_i}$. Then, 
\beqa
M_1 & = & M_{SU(4)} =  
% e^{-i\g_T}\, \frac{t_1}{g^2s + t_1} m_{3/2}
\sqrt3 C m_{3/2} \left[ e^{-i \g_S} \sin \theta \sin^2 \pi\d  +
e^{-i\g_1}  \eta_1 \cos \theta \cos^2\pi\d \right]   \  ,
\nonumber \\[0.1cm]
M_2 & = &  M_{SU(2)_L} = \sqrt3 C m_{3/2} 
e^{-i\g_2}\, \eta_2 \cos \theta   \  ,  
\\ \label{gmasses}
M_3 & = &  M_{SU(2)_R} = \sqrt3 C m_{3/2}
e^{-i\g_3}\, \eta_3 \cos \theta  \  .
\nonumber 
\eeqa

\noindent
\begin{flushleft}
\underline{Scalar masses of SM fields}.
\end{flushleft}

\noindent
The Higgs multiplets appear in a twisted sector (unmagnetized)
{}from  $\D7_2$-$\D7_3$ intersections. One finds then the simple result 
\beq
m^2_{H} = m^2_{3/2} + V_0 - \frac32  C^2 m^2_{3/2}
\left(\sin^2 \theta + \eta_1^2 \cos^2 \theta \right) \ .
\label{smasstu}
\eeq
On the other hand, the chiral quark/lepton fields
appear on  magnetized and twisted sectors $\D7_1$-$\D7_2$ and 
$\D7_1$-$\D7_3$. In particular, all three generations of left-handed
quarks and leptons come from the $\D7_1$-$\D7_2$ sector and
have soft masses
\beqa
m^2_{L_{\a}}  & = &  m^2_{3/2} + V_0 - \frac32 C^2 m^2_{3/2} 
\left(\sin^2 \theta + \eta_3^2 \cos^2 \theta \right)
+  \frac{3}{16\pi^2} C^2 m^2_{3/2}
\sin^2 2\pi\d B_1(\d) |\Theta|^2 
\nonumber \\[0.2cm]
& + & \frac{3}{8\pi} C^2 m^2_{3/2} \sin 2\pi\d \left[ 
2\left(\sin^2 \theta - \eta_1^2 \cos^2\theta  \right)
-\cos 2\pi\d |\Theta|^2 \right] [\log{\frac{u_3}{u_2}} + B_0(\d)] \ ,
\label{smasstm}
\eeqa
where we have defined
\beqa
\Theta & = & e^{-i\g_S} \sin\theta - e^{-i\g_1} \eta_1 \cos\theta   \   ,
\nonumber \\[0.2cm]
B_0(\d) & = & \psi_0(1-\d) - \psi_0(\ts{\oh} -\d) \  ,   \\[0.2cm]   
\label{atajo}
B_1(\d) & = & \psi_1(1-\d) - \psi_1(\ts{\oh} -\d)  \  .
\nonumber
\eeqa
Here $\psi_0(z)= \Gamma^\prime(z)/\Gamma(z)$ and $\psi_1(z)= \psi_0^\prime(z)$.

The three generations of right-handed quarks and leptons
come from the magnetized and twisted $\D7_1$-$\D7_3$ sector. The soft masses 
$m^2_{R_{\a}}$  have the same form as (\ref{smasstm}) except
for the replacements $u_2\leftrightarrow u_3$, $\eta_2\leftrightarrow \eta_3$.
Note that the limit $\d \to 0$ just yields the scalar mass for unmagnetized
twisted fields like the Higgs multiplet, as it should. This also agrees 
with the results obtained for unmagnetized branes in \cite{imr}.

As a check on the results we can use
\beq
\psi_0(\ts{\oh} -\d) = \psi_0(\ts{\oh} +\d) - \pi \tan \pi\d    \  ,
\quad ; \quad \psi_1(\ts{\oh} -\d) = \pi^2 \sec^2 \pi\d - \psi_1(\ts{\oh} +\d)  \  ,
\label{psiprops}
\eeq
to take the limit $\d \to \oh$. This corresponds to infinite magnetic
flux in the 2nd and 3rd torus. In this limit the magnetized $\D7_1$-brane
behaves as a D3-brane. Thus, taking $\d \to \oh$ in (\ref{smasstm}) 
should give the mass squared parameter
of a scalar in a ${\rm D}3$-${\rm D}7_2$ type of sector. In this way
we obtain
\beq
m^2_{C^{37_2}} =  m^2_{3/2} + V_0 - \frac32 C^2 m^2_{3/2}
(1- \eta_2^2) \cos^2 \theta   \  ,
\label{mass37}
\eeq
in accordance with the expected outcome \cite{imr}.

\noindent
\begin{flushleft}
\underline{Trilinear terms of SM fields}.
\end{flushleft}

\noindent
The coupling $H L_{\alpha} R_{\beta}$ is of type  
$C^{7_1 7_2} C^{7_2 7_3} C^{7_3 7_1}$.  
The trilinear term turns out to be
\beq
A_{HLR}  =   \frac{\sqrt3}2 C m_{3/2} \left[ e^{-i\g_S} \sin\theta - 
\sum_i e^{-i\g_i} \eta_i  \cos\theta  
-\frac1{\pi} B_0(\d) \sin 2\pi\d \, \Theta  \right]  \  .
\label{atwi}
\eeq
When $\d \to 0$, the result agrees with that in \cite{imr}.
One can also take the limit when
$\d \to \oh$ which should correspond to a coupling of type 
$C^{3 7_2} C^{7_2 7_3} C^{3 7_3}$. It indeed follows that
\beq
A_{C^{3 7_2} C^{7_2 7_3} C^{3 7_3}} = \frac{\sqrt3}2 C m_{3/2} \left[ 
(e^{-i\g_1} \eta_1 -  e^{-i\g_2} \eta_2 -  e^{-i\g_3} \eta_3) \cos\theta
- e^{-i\g_S} \sin\theta \right] \   ,
\label{tri37}
\eeq
also in agreement with \cite{imr}.

The other relevant trilinear coupling involving SM fields
is of the form $C_1^{7_j} C^{7_2 7_3} C^{7_2 7_3}$, $j=2,3$.
For those couplings one gets trilinear terms $A=-M_j$.
In our case the only such coupling is $XHH$. Then,
\beq
A_{XHH}  = - \sqrt3 C m_{3/2}
e^{-i\g_2}\, \eta_2 \cos \theta   = - M_{SU(2)}  \  .
\label{trili}
\eeq

\noindent
\begin{flushleft}
\underline{Soft terms of non-chiral fields}.
\end{flushleft}

\noindent
Together with the chiral MSSM-like spectrum, there 
are three chiral multiplets in the adjoint of the gauge group
coming from untwisted ${\rm D}7_j$-${\rm D}7_j$ sectors.
We set $\Phi_{ij}=C_i^{7_j}$ and recall that $\Phi_{jj}$
parametrizes the position of each $\D7_j$-brane in transverse space,
whereas the $\Phi_{ij}$ correspond to  Wilson lines
on the two complex dimensions inside the $\D7_j$-brane worldvolume.
For the unmagnetized $\Phi_{i2}$ we find
\beqa
m^2_{12} & = &  m^2_{3/2} + V_0 - 3 C^2 m^2_{3/2} \eta_3^2 \cos^2 \theta   \  , 
\nonumber \\[0.1cm]
m^2_{22} & = & m^2_{3/2} + V_0 - 3 C^2 m^2_{3/2} \sin^2 \theta  \  ,  \\[0.1cm]
\label{smassu}
m^2_{32} & = & m^2_{3/2} + V_0 - 3 C^2 m^2_{3/2} \eta_1^2 \cos^2 \theta  \  .
\nonumber
\eeqa
For the $\Phi_{i3}$ there are analogous results. 
For the magnetized $\Phi_{i1}$ the masses are instead
\beqa
m^2_{11} & = &  m^2_{3/2} + V_0 - 3 C^2 m^2_{3/2} (\sin^2 \theta +  \eta_1^2 \cos^2 \theta)
+ |M_1|^2    \  ,
\nonumber \\[0.1cm]
m^2_{21} & = & m^2_{3/2} + V_0 - 3 C^2 m^2_{3/2} \eta_2^2 \cos^2 \theta  \  ,  \\[0.1cm]
\label{smassum}
m^2_{31} & = & m^2_{3/2} + V_0 - 3 C^2 m^2_{3/2} \eta_3^2 \cos^2 \theta  \  .
\nonumber
\eeqa
In all cases there is a sum rule
\beq
m^2_{1j} + m^2_{2j} + m^2_{3j} = 2V_0  + |M_j|^2   º\  .
\label{sumrule}
\eeq
Besides scalar masses there are also trilinear terms 
associated to the superpotential couplings $\Phi_{1i} \Phi_{2i} \Phi_{3i}$.
They are given by $A_i = -M_i$.

\noindent
\begin{flushleft}
\underline{General structure of soft terms}.
\end{flushleft}

\noindent
Clearly, the structure of soft terms strongly depends on which 
auxiliary fields, either $F_{T_i}$ or $F_S$, dominate 
SUSY-breaking. As a general property one must emphasize that
in all cases the results for scalar masses are 
flavor independent. The trilinear terms involving squarks 
and sleptons are also flavor diagonal. This is an interesting 
property which was not always present in heterotic models and
is welcome in order to suppress too large flavor changing
neutral currents (FCNC). Concerning gaugino masses,
they are equal for the $SU(2)_L\times SU(2)_R$ sector of the theory
but different for the $SU(3+1)$ gauginos which get an extra contribution
proportional to $F_S$ due to the presence of magnetic flux in the 
$\D7_1$-brane. 

One can easily check that in the diluted magnetic flux
limit the results for soft terms agree with those found in 
\cite{imr}. Of particular interest are the extremes in which 
either the overall modulus $T$ or the dilaton $S$ auxiliary
field dominate SUSY-breaking. T-dominance $(\cos \theta = 1)$ 
appears for SUSY-breaking induced by ISD fluxes and we will discuss it 
in detail in the next section. Concerning dilaton
dominance  $(\sin \theta = 1)$, eq. (\ref{smasstm}) shows 
that it is potentially dangerous since squarks
and sleptons typically become tachyonic for small $\d$. 
However, we already mentioned that to stay within the
string perturbative regime the value of $\delta$ cannot be too small, 
c.f. eq.(\ref{gsbound}). Thus, dilaton dominance could still be consistent
if magnetic fluxes are substantial, in fact for $\d \gtrsim 0.258$ when $u_2=u_3$.
One may argue that there are 
scalars whose masses become tachyonic in the limit $\sin \theta =1$ and have 
no dependence on $\delta$, since they are related to unmagnetized branes.
This is the case of the Higgs and the 
non-chiral scalars $\Phi_{22}$ and $\Phi_{33}$
parameterizing the position of branes $\D7_2$, $\D7_3$. However, for both
there could be extra contributions to their masses which would render them
non-tachyonic. In the case of the Higgs multiplet we already mentioned
that they get an additional SUSY mass term for $\langle X \rangle\not=0$. Concerning
the $\Phi_{22}$, $\Phi_{33}$ scalars, they may also have SUSY
$\mu $-terms. For example, in flux induced SUSY-breaking, 
fluxes of type $(2,1)$ or $(1,2)$ could give rise to
such terms (see \cite{ciu2}). All in all, whereas T-dominance 
always leads to a non-tachyonic structure of soft terms, in the case of
dilaton dominance substantial magnetic fluxes and additional positive
contribution for Higgsses and some adjoints are required to avoid tachyons.

\subsection{Soft terms induced by fluxes}

We now wish to discuss the situation in which the moduli superpotential 
is just given by $W_f$, c.f. (\ref{wflux}). Then, fluxes are the only 
source of SUSY breaking.

\noindent
\begin{flushleft}
\underline{Soft terms from T-dominance (ISD fluxes)}.
\end{flushleft}

\noindent
A concrete realization of T-dominance arises when the flux $G_3$ is generic 
ISD. In this case, $\langle F^S \rangle = \langle F^{U_i} \rangle= 0$
and only $\langle F^{T_i} \rangle \not= 0$. The cosmological constant
vanishes automatically since $W_f$ is independent
of $T_i$ and $\hat K$ is of no-scale form. Thus, $V_0=0$ and the other
relevant vevs are
\beq
m^2_{3/2} = \frac{|W_f|^2}{P}   \quad ; \quad
\bar F^{\bar T_i} = -\frac{t_i W_f}{P^{1/2}} = e^{i\g_T}\,  t_i \, m_{3/2}
 \  ,
\label{isd}
\eeq
where $P=s \prod_i t_i u_i$. Comparing with (\ref{vevstino}) and  (\ref{cconst})
shows that $C=1$, $\cos \theta=1$, $\eta_i=1/\sqrt3$, $\g_i=\g_T$, $\forall i$.
The soft terms follow substituting these values in the general expressions.
The results are collected in Table 2 \footnote{When the ISD flux has both (2,1) and (0,3)
components there is an induced supersymmetric $\mu$ term and an extra soft bilinear parameter
for the $\Phi_{ii}$ scalars \cite{ciu2}.}.

\begin{table}[htb]
\renewcommand{\arraystretch}{1.25}
\begin{center}
\begin{tabular}{|c|c|}
\hline 
$m^2_{L_\a}$  &  $\frac12 \ -
\frac1{8\pi} \sin 2\pi\d (2 + \cos 2\pi\d)[\log{\frac{u_3}{u_2}} + B_0(\d)]
+ \frac1{16\pi^2} \sin^2 2\pi\d B_1(\d)$ \\
\hline
$m^2_{R_\b}$  &  $\frac12 \ -
\frac1{8\pi} \sin 2\pi\d (2 + \cos 2\pi\d)[\log{\frac{u_2}{u_3}} + B_0(\d)]
+ \frac1{16\pi^2} \sin^2 2\pi\d B_1(\d)$ \\
\hline
$m^2_{H} $  &   $ \frac {1}{2} $     \\
\hline \hline
$ M_{SU(3+1)}$  &   $e^{-i\gamma _T} \cos^2\pi \delta $  \\
\hline
$ M_{SU(2)_L}$  &   $e^{-i\gamma _T}$     \\
\hline
 $ M_{SU(2)_R}$  &   $e^{-i\gamma _T} $    \\
\hline \hline
 $A_{HL_\a R_\b}$  &   $e^{-i\gamma _T}\ \left[ -\frac32 \ +
\frac1{2\pi} \sin 2\pi\d B_0(\d) \right]$ \\
\hline
$A_{XHH}$  &   $ -e^{-i\gamma _T}$   \\
\hline\hline
$m^2_{\Phi_{jj}}$ & $|M_j|^2$ \\
\hline
$m^2_{\Phi_{ij}}$ & 0 \\
\hline
\end{tabular}
\caption{Soft terms for 
T-dominant ISD fluxes. Results are given in $m_{3/2}$ units.}
\end{center}
\label{resuT}
\end{table}

It is straightforward to expand the soft terms near $\d=0$.
For example,
\beqa
m^2_{L_\a}  & = &   m^2_{3/2} \left(\frac12  - \frac34 \d \log \frac{4u_3}{u_2} -
\frac{\pi^2}3 \d^2 + \cdots \right) \ , 
\nonumber \\[0.1cm]
A_{HL_\a R_\b} & = & e^{-i\g_T}\, m_{3/2}
\left(-\frac32 + 2 \d \log 2 + \frac{\pi^2}3 \d^2 + \cdots \right)  \  .
\label{softexp}
\eeqa
Using (\ref{psiprops}) we can also take
the limit $\d \to \oh$ in which $\D7_1 \to \D_3$.
We find, $m^2_{L_\a} \to 0$ and $A_{HL_\a R_\b} \to -\frac12 e^{-i\g_T}\, m_{3/2}$,
matching results of \cite{imr}.
 
Note that the  structure of soft terms in this subsection corresponds to
SUSY-breaking induced by the auxiliary field of the overall 
K\"ahler modulus $T$, and the fact that it may be induced by fluxes 
plays no role in the obtained results.
A few comments on the structure of soft terms in this simple case are
in order.

\begin{itemize}

\item 
The structure of scalar soft terms is not universal, i.e. 
the scalar masses of left-handed sfermions, right-handed sfermions
and Higgsses are different. However they are {\it flavor
independent}. This is due to the origin of family replication 
in this class of models. The massless chiral fermions come
in identical replicas with diagonal kinetic terms.
Concerning gaugino masses, as we said the $SU(2)_L$ and
$SU(2)_R$ gaugino masses are identical but that
of $SU(3+1)$ is different.

\item 
Note that if at the end of the day a non-vanishing
FI-term (\ref{fi}) is present, an extra contribution to 
squark/slepton masses with opposite signs for left and
right-handed fields will be added. This contribution will
not be present for the Higgs fields which are neutral 
under the anomalous $U(1)$.

\item 
In the formal limit $\delta \rightarrow 0$ corresponding 
to diluted magnetic fluxes one obtains particularly simple and universal results
for soft terms :
\beqa
m^2_{L_\a} & = & m^2_{R_\a} \ =\  m^2_{H}\  =\ \frac12 m_{3/2}^2   \\
\nonumber
 M_{SU(3+1)} & = & M_{SU(2)_L} \ =\ M_{SU(2)_R} \ =\ 
e^{-i\gamma _T} \ m_{3/2} \\ 
\nonumber
A_{HL_\a R_\b}  & = & -\frac {3}{2} m_{3/2} e^{-i\gamma _T}\  \\
A_{XHH} & = &  e^{-i\gamma _T} \ m_{3/2}
\label{softguay}
\eeqa
These results correspond to those advanced in section 6 of \cite{iflux}
for $\mu = \langle X \rangle$ and $\xi =1/2$.

\end{itemize}

\noindent
\begin{flushleft}
\underline{IASD fluxes and dilaton dominance}.
\end{flushleft}

\noindent
Let us now analyze the mass parameters generated 
by the presence of IASD fluxes.
Some words of caution should first be given.
Care should be taken in comparing the results below for IASD fluxes to
those for dilaton dominance ($\sin \theta =1$) in
the previous section. Indeed in the analysis of that section
the values of $V_0$, $\sin \theta$, $m_{3/2}$, appear as independent
parameters. Thus, in principle one can conceive a situation with
$\sin \theta =1$, $V_0=0$ and $m_{3/2}\not =0$ with SUSY broken
in Minkowski space. However, with IASD (3,0) fluxes 
one can show that $V_0\not = 0$ and $m_{3/2}=0$, so we have broken
SUSY in de Sitter space. 
We will still provide the generated mass terms for completeness.

If we consider $W_f$  as our only source of SUSY-breaking
dilaton dominance appears 
when the flux  $G_3$ is IASD of (3,0) type. In 
the absence of a $W_{np}$ term this background is not in general a solution
of the Type IIB equations of motion. 
However, it is a simple example of S-dominance because
$\langle F^{T_i} \rangle = \langle F^{U_i} \rangle= 0$ but 
$\langle F^S \rangle \not= 0$. In this case $m_{3/2} = 0$ automatically
and, as we said, there is a cosmological constant 
$V_0\not= 0$. In terms of $Y_f=\int {\bar G_3} \wedge
\Omega$
one finds 
\beq
V_0 = \frac{|Y_f|^2}{P}   \quad ; \quad
\bar F^{\bar S} = -\frac{s Y_f}{P^{1/2}} = e^{i\g_S}\,  s \, \sqrt{V_0}
 \  .
\label{iasd}
\eeq
Hence, the soft terms can be obtained from the general expressions setting
$\sin \theta = 1$ and $C m_{3/2} \to \sqrt{V_0/3}$. 
Results are displayed in Table 3.
Note that the mass parameters are in general not tachyonic.
This is not in contradiction with our results discussed at the end
of section 3, since there we assumed arbitrary  $V_0$ and $m_{3/2}\not= 0$,
whereas IASD $(3,0)$ fluxes lead to $V_0\not=0$ and $m_{3/2}=0$.

\begin{table}[htb]
\renewcommand{\arraystretch}{1.25}
\begin{center}
\begin{tabular}{|c|c|}
\hline
$m^2_{L_\a}$  &  $\frac12 \ +
\frac1{8\pi} \sin 2\pi\d (2 - \cos 2\pi\d)[\log{\frac{u_3}{u_2}} + B_0(\d)]
+ \frac1{16\pi^2} \sin^2 2\pi\d B_1(\d)$ \\
\hline
$m^2_{R_\b}$  &   $\frac12 \ +
\frac1{8\pi} \sin 2\pi\d (2 - \cos 2\pi\d)[\log{\frac{u_2}{u_3}} + B_0(\d)]
+ \frac1{16\pi^2} \sin^2 2\pi\d B_1(\d)$ \\
\hline
$m^2_{H} $  &   $ \frac {1}{2} $     \\
\hline \hline
$ M_{SU(3+1)}$  &   $e^{-i\gamma _S} \sin^2\pi \delta $  \\
\hline
$ M_{SU(2)_L}$  &   0     \\
\hline
 $ M_{SU(2)_R}$  &  0   \\
\hline \hline
 $A_{HL_\a R_\b}$  &   $e^{-i\gamma _S}\ \left[ \frac12 \ -
\frac1{2\pi} \sin 2\pi\d B_0(\d) \right]$ \\
\hline
$A_{XHH}$  &  0   \\
\hline\hline
$m^2_{\Phi_{jj}}$ & $|M_j|^2$ \\
\hline
$m^2_{\Phi_{ij}}$ & 1 \\
\hline
\end{tabular}
\caption{Soft terms for
S-dominant IASD fluxes. Results are given in $V_0$ units.}
\end{center}
\label{resuS}
\end{table}

When ISD and IASD fluxes are turned on simultaneously the auxiliary fields,
$m_{3/2}$ and $V_0$ are just given by (\ref{isd}) and (\ref{iasd}). Then,  
$\eta_i=1/\sqrt3$, $\g_i=\g_T$, $\forall i$, $3 \tan^2 \theta=V_0/3m_{3/2}$,
and $C=\sec \theta$. The soft terms can be found 
substituting these values in the general expressions.
In most cases it suffices to add the entries in Tables 2 and 3. 

%\vspace*{1.0cm}

Let us end this section with some comments concerning 
the mass scales in this model. Note that as it stands,
in a {\it toroidal } model like this, the string scale $M_s$
should be of order (or slightly smaller) than the Planck scale.
Indeed, both scales are related by eq.(\ref{mplanck}).
Although one may think that one can make $M_{Pl}>>M_s$ by
taking $t_i$ very large, that would make the SM gauge
couplings  unacceptably small, as shown by eqs.(\ref{efes}).
On the other hand, the size of SUSY-breaking soft terms 
depends on the value of the gravitino mass in these theories.
In general, if the source of SUSY-breaking is not specified,
as in section 3, one can assume that $m_{3/2}$ may be small,
i.e. of order the electroweak scale, as in the canonical
approach to gravity mediated SUSY-breaking models.
This was our general philosophy in the first part of section 4.
On the other hand, if one insists that the microscopic source of
SUSY-breaking is some ISD flux in a {\it toroidal } setting , then the 
gravitino mass is given by eq.(\ref{isd}). In that case, since the 
$t_i$ fields cannot be too large, the gravitino mass is 
of order the string scale (which is only slightly smaller 
than $M_{Pl}$) and hence too large to lead to a solution of the
hierarchy problem. This is the fact already mentioned
in the introduction. However, as we said, this is
a particular property of toroidal settings in which fluxes 
are distributed uniformly in extra dimensions.  One can 
conceive an embedding of the MSSM-like brane setting 
in \cite{cim1} into a CY/F-theory compactification in
which the distribution of fluxes in extra dimensions is not
constant and hierarchically small soft terms may appear.

\section{Conclusions}
\label{sec:fin}

In this paper we have computed the SUSY-breaking 
soft terms for the MSSM-like model introduced in \cite{cim1}
under the assumption of  generic vevs for the auxiliary fields $F_{T_i}$ and $F_S$.
We provide the soft terms as explicit functions
of the gravitino mass, goldstino angle
and a  parameter $\delta $ that characterizes the magnetic flux
in one of the brane stacks. We find that the case of 
isotropic $T$-dominance is particularly interesting since 
it always leads to simple results with no tachyons. 
For dilaton dominance there is the risk of getting
some tachyonic masses for SM fields unless magnetic fluxes 
are large and additional sources for masses of non-chiral fields are present.
The case of isotropic $T$-dominance appears in particular  
when SUSY-breaking is triggered by ISD antisymmetric Type IIB 
fluxes. We argue that although in a toroidal setting the 
soft terms induced by fluxes are typically too large,
they may be hierarchically small in more general CY/F-theory
embeddings of this MSSM-like brane configuration. 

The results for soft terms in $T$-dominance
are summarized in Table 2, and take an
even simpler form (\ref{softguay}) in the dilute flux limit $\delta \rightarrow
0$. They  are flavor universal and depend only 
on the values of $m_{3/2}$, $\delta $ and 
a complex phase $\gamma_T$.

%\vspace*{0.5cm}

\newpage

{\bf \large Acknowledgments}

We  thank P.G. C\'amara, D. Cremades,   F. Marchesano,
F. Quevedo, S. Theisen, and A. Uranga for useful  discussions.
This work has been partially supported by the European Commission under
the RTN European Program  MRTN-CT-2004-503369 and the CICYT (Spain).

%\newpage

%\vspace*{0.1cm}  

\end{document}